\documentstyle[epsfig,11pt]{article}
\title{\bf Triple-pomeron amplitude in the effective action approach}
\author{M.A.Braun\\
{\it S.Petersburg State University, Russia}}
\begin{document}

\maketitle
\input epsf

\def\beq{\begin{equation}}
\def\eeq{\end{equation}}
\def\disc{{\rm Disc}}
\def\lra{\leftrightarrow}
\def\im{{\rm Im}\,}
\def\re{{\rm Re}\,}
\def\lra{\leftrightarrow}
\def\tm{m}
\def\tp{p}
\def\qa{q_{1+}}
\def\qb{q_{2+}}
\def\qc{q_{3+}}
\def\ra{r_{1-}}
\def\rb{r_{2-}}
\def\rc{r_{3-}}
\def\rd{r_{4-}}
\def\pta{p_{1\perp}^2}
\def\ptb{p_{2\perp}^2}
\def\ptc{p_{3\perp}^2}
\def\pa{p_{1-}}
\def\pb{p_{2-}}
\def\pc{p_{3-}}
\def\ppe{p_{2-}-p_{1-}-e_-}
\def\ts{\tilde{s}}
\def\e{e_-}
\def\sig{{\rm sign}\,(q_{1+})}
\def\ep{e_-}
\def\pd{\partial}
\def\tp{\tilde{p}}
\def\bal{\bar{\alpha}_s}
\def\br{{\bf r}}

\begin{abstract}
{In the effective action approach the imaginary part of the triple pomeron amplitude  is calculated.
The found dependence on the longitudinal momentum transfer $\e$ is found to separate as a simple factor $1/|\e|$.
This result is used to calculate the high-mass diffraction on a hadron and double scattering cross-section off a composite target}
\end{abstract}

\section {Motivation}
\begin{figure}
\begin{center}
\epsfig{file=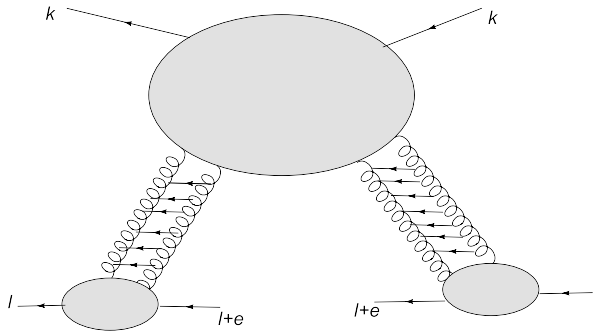, width=8 cm}
\caption{Diffractive scattering off a hadron}
\label{fig1}
\end{center}
\end{figure}

\begin{figure}
\begin{center}
\epsfig{file=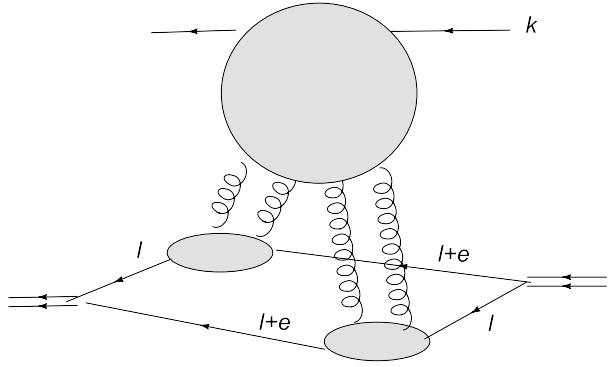, width=8 cm}
\caption{Double scattering on the deuteron or nucleus. In the second case
the spectator nucleons are not shown}
\label{fig2}
\end{center}
\end{figure}

Rather long ago in the study of interaction of a colorless projectile with two colorless targets the triple-pomeron vertex $\Gamma$
was constructed both in the BFKL approach ~\cite{bartels,barwue} and
dipole picture ~\cite{mueller}.
The simplest scattering amplitudes involving the triple pomeron vertex include the diffractive scattering on the hadron
Fig. \ref{fig1} and the double scattering on the deuteron (or nucleus) Fig. \ref{fig2}.

Consider first the diffractive scattering Fig. \ref{fig1}.
 We denote the scattering amplitude
 as $H$, the momentum of the projectile in the c.m. system as $k$ with $k_-=k_\perp=0$ and the initial and final momenta of each of the target
as $l$ and $l'$ With $l_+=l_\perp=0$ and $l'-l=e$.  The diffractive mass squared is  $M^2=(k+l-l')^2=-2k_+\e$
and we assume $s>>M^2>> {l'_\perp}^2$
The diffractive cross-section is given by
\beq
d\sigma=\frac{d\e d^2l'_\perp}{16\pi^3l_-}\frac{1}{s}\,\frac{1}{2i}{\rm Disc}_d{\cal A}(e),
\label{sigd}
\eeq
where ${\rm Disc}_d{\cal A}$ is a particular, diffractive discontinuity of the amplitude on the cut passing between the two target in Fig. \ref{fig1}.
Using the relation between $e_-$ and the diffractive mass we can rewrite (\ref{sigd}) as
\beq
d\sigma=\frac{dM^2d^2l'_\perp}{16\pi^3 s^2}\,\frac{1}{2i}{\rm Disc}_d {\cal A}(e).
\label{eqa1}
\eeq
It is convenient to separate the trivial energetic factor and define
\beq
\frac{1}{i}{\rm Disc}_d{\cal A}=\frac{2s^2}{k_+}H.
\label{defh}
\eeq
In terms of amplitude $H$ we find
\beq
d\sigma=\frac{dM^2d^2l'_\perp}{16\pi^3 k_+}H(e)=\frac{dM^2d^2l'_\perp}{8\pi^3M^2}|\e| H(e).
\label{eqa2}
\eeq

Passing to the double scattering Fig. \ref{fig2} and using expressions from ~\cite{bra5}
we have the cross-sections on the nucleus
\beq
\frac{d\sigma_A}{d^2b}=\frac{A(A-1)}{4\pi k_+s}T^2(b)\int d\e\im {\cal A}(e)
\label{crseca}
\eeq
and on the deuteron
\beq
\sigma_d=\frac{1}{4\pi^2 k_+s} \Big<\frac{1}{r^2}\Big>_d\int d\e\im {\cal A}(\e).
\label{crsecd}
\eeq
where this time $e_\perp=0$ and
 one has to take the total imaginary part of ${\cal A}$. However it is trivially related to its
diffractive part, namely (~\cite{agk})
\beq
\im {\cal A}(e)=-\frac{1}{2i}{\rm Disc}_d {\cal A}(e).
\label{agk}
\eeq
In fact this fact trivially follows from the relation between the uncut and cut pomerons $P$
With real $P$ the uncut pomeron is $-P$ and the cut one $2P$. So in Fig. \ref{fig1} we have 2 uncut outgoing pomerons
and in Fig/ \ref{fig2} we have either 2 uncut pomerons or 2 cut pomerons or 4 pairs cut+uncut pomerons giving the total number of pomeron pairs
1:2:-4. This leads to relation (\ref{agk})
Relating ${\cal A}$ and $H$ by (\ref{defh}) we find finally
for the nucleus
\beq
\frac{d\sigma_A}{d^2b}=-\frac{A(A-1)}{2\pi}T^2(b)\int d\e H(\e)
\label{crseca1}
\eeq
and for the deuteron
\beq
\sigma_d=-\frac{1}{\pi^2} \Big<\frac{1}{r^2}\Big>_d\int d\e H(\e).
\label{crsecd1}
\eeq

Inspecting expressions for the cross-sections for both cases one arrives at the following conclusions.

First it is sufficient to know the diffractive cross-sections. The cross-sections for the double scattering
can  be found from the AGK relation (\ref{agk}) after attaching the necessary factor for the compound target.

\begin{figure}
\begin{center}
\epsfig{file=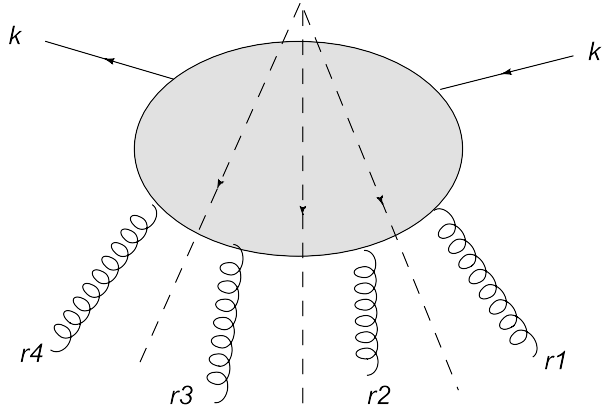, width=8 cm}
\caption{Triple cut of the triple-pomeron amplitude}
\label{fig3}
\end{center}
\end{figure}

Second one has to know the dependence of the amplitude $H$ on the longitudinal momentum transfer $\e$. This can be done
only if one retains the dependence on longitudinal variable in the calculation of the relevant diagrams. In the old derivations
one studied the diagrams with free outgoing reggeons  and calculated the triple discontinuity on the three cuts passing between them
Fig. \ref{fig3}. After that one integrated over the three "-" components of these  reggeons to obtain a purely transversal
expression (see e.g. ~\cite{bartels, barew, hent}) Passage to the final pomerons was then made directly in the transverse space.
The $\e$ dependence was traded for the $M^2$ dependence
corresponding to the rapidity of the incoming reggeon. Apart from the somewhat dubious validity of this procedure  for the diffractive scattering
it is not applicable to the double scattering where one has to integrate over all values of $\e$

The clear way to overcome this difficulty is to use Lipatov's effective action approach (LEA), which generates amplitudes with full dependence
on the longitudinal variables. This motivates our study, in which we calculate the amplitude ${\cal A}(e)$ using the effective action.
In fact the amplitude ${\cal A}$ itself is most complicated and has to be calculated in the general gauge due to the fact that all
intermediate gluons initially lie off the mass shall. However the amplitude ${\cal A}$ as a whole appears only in the expression for the
elastic scattering on the composite target where one should know $|{\cal A}|^2$ at the momentum transfer
different from zero. Leaving this task for future calculations we restrict here to the discussed physical cases where one only needs
the diffractive imaginary part of ${\cal A}$. Apart from drastically reducing the number of relevant diagrams it allows to work in the
light-cone gauge for the intermediate real gluon taking its polarization vector $\epsilon$ orthogonal to $l$, that is with $\epsilon_+=0$.

To see the $\e$ dependence it is sufficient to use the perturbative approach and start with the lowest order. This means that
we can approximate all pomerons with the double gluon exchange. Also one can take simple $q\bar{q}$ loops for the participants.
Our calculations are divided in parts corresponding to the number of incoming reggeons
attached to the projectile 2,3 and 4  and studied in the next sections 2,3 and 4 respectively.
Having in mind the applications either to the diffractive scattering or double scattering on the nucleus,
we consider the case when the incoming pomeron is the forward one
but the two outgoing ones  are generally non-forward.

%%%%%%%%%%%%%%%%%%%%%%%%%%%%%%%%%%%%%%%%%%%%%%%%%%%%%%%%%%%%%%
%%%%%%%%%%%%%%%%%%%%%%%%%%%%%%%%%%%%%%%%%%%%%%%%%%%%%%%%%%%%%
%%%%%%%%%%%%%%%%%%%%%%%%%%%%%%%%%%%%%%%%%%%%%%%%%%%%%%%%%%%%%%%%%
\subsection{The relevant vertices in the effective action}
%%%%%%%%%%%%%%%%%%%%%%%%%%%%%%%%%%%%%%%%%%%%%%%%%%%%%%%%%
%%%%%%%%%%%%%%%%%%% FROM OUR PAPER W LIPATOV %%%%%%%%%%%%%%%%%%%
The LEA approach was presented in the detail in ~\cite{lipatov} and the following Fenian rules were
formulated in ~\cite{antonov}.  Here we only briefly discuss its main points to  make the following derivation.
In the LEA  within a rapidity slice of finite dimension gluons
are described by the usual (matrix) gluon field $G=-it^{a}G^{a}$.
The two reggeon field $R_{\pm}$ with the only non-zero longitudinal components connect slices with widely different rancidities.
The effective Lagrangian describes the interaction of gluons and reggeons with a given rapidity slice.It takes the form \cite{lipatov}:
\[
{\cal L}_{eff}={\cal L}_{QCD}(V+G)\]\beq
+ 2 {\rm Tar}\,Big\{\Big(j_+(G+R)-R\Big)\pd^2_{\perp} G_-
+ \Big(j_-(G+R)-R\Big)\pd^2_{\perp} G_+\Big\}),
\label{e1}
\eeq
where ${\cal L}_{QCD}(V)$ is the usual QCD Lagrangian and
\beq
j_{\pm}(G)=
\sum_{n=0}^{\infty}(-g)^nG_{\pm}(\pd_\pm^{-1}G_\pm)^n.
\label{e2}
\eeq
The shift $G\to G+R$ with $R_\perp=0$ is done
to exclude direct gluon-reggeon transitions.
The reggeon propagator in momentum representation is
\beq
\Delta^{ab}(y'-y,q)=<R_+^a(y')R_-^b(y)>=-i\frac{\delta_{ab}}{q_\perp^2}
\,\theta(y'-y).
\label{e3}
\eeq
Here $a,b$ are color indices.
It couples  field $R_-$ interacting with a group of a higher rapidity $y'$
and  field $R_+$ interacting with a group of a smaller rapidity $y$.
From the kinematical constraints it  follows that
\beq
\partial_{\mp} R_{\pm}=0.
\label{e4}
\eeq
The effective Lagrangian generates elementary vertices for the interaction of gluons and reggeons which come both from the QCD
part and the rest "induced" part and so separate  in basic and induced vertices. In higher orders propagation of
intermediate virtual gluons gives rise to more complicated compound vertices, which contain one or several off-shell intermediate gluon states.

In our calculations it will be sufficient to know two basic vertices for gluon production in the interaction of one incoming reggeon
with one outgoing ("the Lipatov vertex") $\Gamma_{R\to G+R}$ or two outgoing reggeons $\Gamma_{R\to G+2R}$. The first vertex has been known
since long ago (see ~\cite{bfkl}), the second was calculated in LEA in ~\cite{bravyaz}. We reproduce them here in the light-cone gauge
with respect to the target with the gluon polarization vector $\epsilon$ orthogonal to the incoming momentum of the target $l$ : $(l\epsilon)=0$,
so that $\epsilon_+=0$.
\begin{figure}
\begin{center}
\epsfig{file=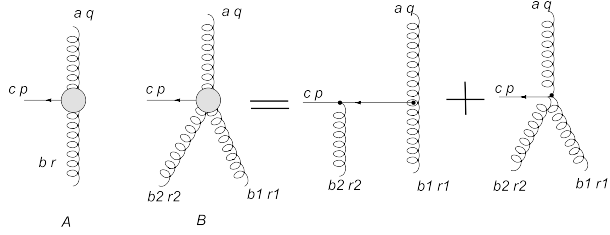, width=8 cm}
\caption{Vertexes $\Gamma_{R\to G+R}$ (A) and $\Gamma_{R\to G+2R}$ (B).
The second vertex should be symmetrized in the outgoing reggeons}
\label{fig4}
\end{center}
\end{figure}

In this gauge the Lipatov vertex is (Fig. \ref{fig4},A)
\beq
\Gamma_{R\to G+R}=g q^2L(p,r),\ \ L(p,q)=\frac{(pe)_\perp}{p_\perp^2}-\frac{(p+r,e)_\perp}{(p+r)_\perp^2}.
\label{lv}
\eeq
Vertex $\Gamma_{R\to G+2R}$ is given by the sum ( Fig. \ref{fig4},B)
\beq
\Gamma_{R\to G+2R}=W_1+R_1+(1\lra 2).
\label{rgrr}
\eeq
Here
vertex $R_1$ is
\beq
R_1=ig^2\frac{q_\perp^2}{\ra}f^{ab_1d}f^{db_2c}L(p,r_2).
\label{r1}
\eeq
Vertex $W_1$ is
\beq
W_1=-ig^2\frac{2q_+q_\perp^2}{(q-r_1)^2+i0}f^{ab_1d}f^{db_2c}B(p,r_2,r_1)
\label{w1}
\eeq
where the "Barters vertex" $B$ is
\beq
B(p,r_2,r_1)=L(p+r_2,r_1).
\label{bv}
\eeq
$(1\lra 2)$ means the interchange of the two outgoing reggeons.
Note that $R_1$ contains a singularity at $\ra=0$, which should be understood in the principal value
prescription.

\section {Two reggeons attached to the projectile}
\begin{figure}
\begin{center}
\epsfig{file=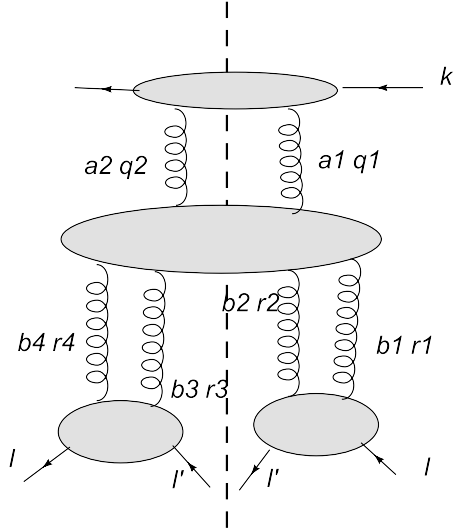, width=6 cm}
\caption{The diffractive amplitude with two reggeons attached to the projectile}
\label{fig5}
\end{center}
\end{figure}

The diagram corresponding to this amplitude is shown in Fig. \ref{fig5}. For the diffractive contribution the cut should go
between the two targets.
In the lowest order the blobs are usually taken as $q\bar{q}$ loops with the two reggeons attached to the
quarks and antiquarks in all different ways.  The reggeons are just gluons with the propagators depending on only their transverse
momenta.
The projectile impact factor depends on $q_{1-}$. The target impact factors depend on "+" components of $r$'s.
The central blob is the vertex $\Gamma_{2R\to 4R}$ for the transition of  two incoming reggeons into four outgoing ones. Apart from the
transverse momenta it depends on $\qa=-\qb$, $\ra$ and $\rd$. One has $r_1+r_2=-r_3-r_4=e$ with $e_+=0$

Integration over $q_-$, $r_{1+}$ and $r_{4+}$ are factored out. Only the impact factors depend on them. So one can immediately
do these integrations by using energetic variables $(k-q_1)^2$, $(l+r_1)^2$ and $(l+r_4)^2$ and rotating the Feynman integration contour
around the cut of the impact factors. In this way one obtains the standard Impact functions $D^{(0)}$ depending only on
the corresponding transverse momenta of the attached reggeons multiplied by the overall energetic factor   $2s^2/k_+$ entering
 (\ref{defh}). Factors $1/2\pi$ associated with these longitudinal integrations are usually included into the
integration volume $d\tau_\perp$ over the transverse momenta in which for each momentum $q$ appears $d^2q/8\pi^3$.
Also since the singularities of the impact factors are supposed to lie at limited values the momenta,
$q_-$, $r_{1+}$ and $r_{4+}$ become of the order $1/k_+$ that is practically zero at large energies.
As a result the incoming reggeons acquire $q_-=0$ and the outgoing ones $r_+=0$ in accordance with the effective action.

In this way one obtains the diffractive amplitude $H_2$ coming from the contributions with two reggeons attached to the projectile
in the form of the integral
\beq
H_2(e)=\int d\tau_\perp P^{(0)}(q_1,q_2)P^{(0)}(r_1,r_2)P^{(0)}(r_4,r_3)\int\frac{d\qa d\ra d\rd}{(2\pi)^3}\Gamma_{2R\to 4R}(\qa,\ra,\rb).
\label{d2}
\eeq
Here $d\tau_\perp$ is the phase volume for the integration over the transverse momenta $q_1$,$r_1$ and $r_4$ and
it is implied that the vertex $\Gamma_{2R\to 4R}$ also depends on these momenta.

The summation over colors gives $N^4$.

The integration over $\qa$ is done due to the cut which provides $-2\pi\delta\Big((q_1-e)^2\Big)$. One gets factor $-i/4|\e|$
and puts $\qa=p_\perp^2/2\e=-\qb $.  Note that $p_-=-\e$ has to be greater than zero, which automatically requires $\e<0$.
Otherwise the discontinuity is zero. With $\e<0$ one has $\qa>0$ and $\qb<0$.

The rest part of $\Gamma_{2R\to 4R}(\qa,\ra,\rb)$ is the product of two vertices $\Gamma_{R\to G+2R}$ given by (\ref{rgrr}).
Integration over $\ra$ and $\rd$ are separated. One gets
\beq
\int\frac{d\ra}{2\pi}\Gamma_{R\to G+2R}(r_1,r_2|q_1)=2\int\frac{d\ra}{2\pi}W(r_1,r_2|q_1)
\eeq
because the interchange $(1\lra 2)$ does not change the result and integration over of $R$ gives zero
due to the principal value prescription.
We have
\[
\int\frac{d\ra}{2\pi}\frac{1}{(q_1-r_1)^2+i0}= \int\frac{d\ra}{2\pi}\frac{1}{-2\qa\ra+(q_1-r_1)_\perp^2+i0}=-i\frac{1}{4|\qa|},\]
so that after integration we get
\[\int\frac{d\ra}{2\pi}W(r_1,r_2|q_1)=-g^2\frac{\qa q_1^2}{2|\qa|}B(p,r_2,r_1),\]
where $p=q_1-e$.
Integration of the second vertex
 $W$ gives in the same way
\[\int\frac{d\rd}{2\pi}W(r_4,r_3|q_2)=-g^2\frac{\qb q_2^2}{2|\qb|}B(-p,r_3,r_4).\]
Conjugation requires to invert all vectors but this does not change the  $\qb B(-p,r_3,r_4)$.
The emerging
factor  $\qa\qb/|\qa||\qb|=-1$.

Collecting all factors  we  find
\[
H_2(e)=\frac{1}{2|\e|}g^4N^4\int d\tau_\perp D^{(0)}(q_1,q_2)P^{(0)}(r_1,r_2)P^{(0)}(r_4,r_3)\]\beq\times
\Big(\frac{p+r_2}{(p+r_2)^2}-\frac{q_1}{q_1^2}\Big)\Big(\frac{-p+r_3}{(-p+r_3)^2}-\frac{q_2}{q_2^2}\Big),
\label{d21}
\eeq
where  one can consider all vectors to be Euclidean 2-dimensional.
The product of two vectors which appears is in fact the well-known Bartels kernel for transition of two reggeons into four
\beq
K_{2\to 4}(r_4,r_3,r_2,r_1|q_2,q_1)=K_{2\to 3}(r_4,r_3+r_2,r_1q_2,q_1).
\label{k24}
\eeq
Indeed  denote $p_1=p+r_2$ and $p_3=-p+r_3$ then we have the product in (\ref{d21})
\[\Big(\frac{p_1}{p_1^2}-\frac{q_1}{q_1^2}\Big)\Big(\frac{p_3}{p_3^2}-\frac{q_2}{q_2^2}\Big)=
\frac{p_1p_3}{p_1^2p_3^2}+\frac{q_1q_2}{q_1^2q_2^2}-\frac{p_1q_2}{p_1^2q_2^2}-
\frac{p_3 q_1}{p_3^2q_1^2}\]\[=
\frac{1}{2p_1p_3}\Big((p_1+p_3)^2-p_1^2-p_3^2\Big)+\frac{1}{2q_1^2q_2^2}\Big((q_1+q_2)^2-q_1^2-q_2^2\Big)\]\[-
\frac{1}{2p_1^2q_2^2}\Big((p_1+q_2)^2)-p_1^2-q_2^2\Big)-\frac{1}{2p_3^2q_1^2}\Big((p_3+q_1)^2-p_3^2-q_1^2\Big)\]\[=
\frac{(p_1+p_3)^2}{2p_{1}^2p_{3}^2}+\frac{(q_1+q_2)^2}{2q_1^2q_2^2}
-\frac{(p_1+q_2)^2}{2p_1^2q_2^2}-\frac{p_3+q_1)^2}{2p_3^2q_1^2}\]\beq
=\frac{1}{2}K_{2\to 4}(r_4,r_3+r_2,r_1|q_2,q_1).
\label{derivk}
\eeq
Here all vectors are here Euclidean 2-dimensional. In the last expression in (\ref{derivk}) we take into account
that $p_1+p_3=r_2+r_3$.

So we finally find
\beq
H_2(e)=\frac{1}{4|\e|}g^4N^4\int d\tau_\perp D^{(0)}(q_1,q_2)P^{(0)}r_1,r_2)P^{(0)}(r_4r_3)
K_{2\to 4}(r_4,r_3,r_2,r_1|q_2,q_1)
\label{d22}
\eeq
with $q_1+q_2=0$ and $r_1+r_2=-r_3-r_4=e$.
This is the same expression which  one obtains using the dispersive multiple cut approach and  passing to the transverse space,
except for the new factor $1/|\e|$.
%%%%%%%%%%%%%%%%%%%%%%%% Made up to this point%%%%%%%%%%%%%%%%%%%%%%

\section{Three reggeons attached to the projectile}
\begin{figure}
\begin{center}
\epsfig{file=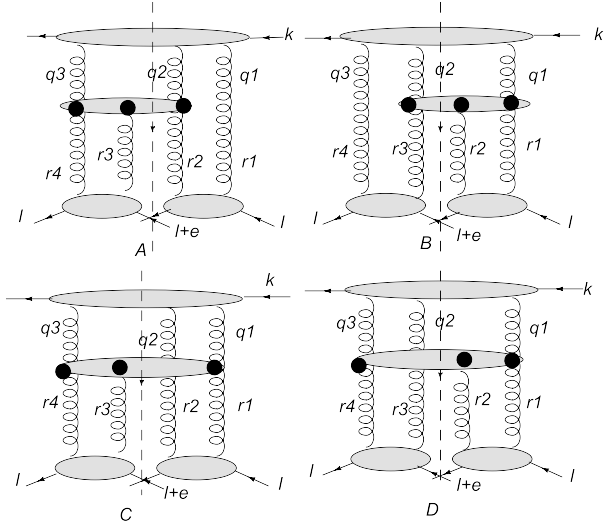, width=8 cm}
\caption{The diffractive amplitudes with three reggeons attached to the projectile}
\label{fig6}
\end{center}
\end{figure}

Four diagrams corresponding to this amplitude are shown in Fig. \ref{fig6}. For the diffractive contribution the cut should go
in between the two targets.
In the lowest order of perturbations the blobs can be taken as $q\bar{q}$ loops with three reggeon attached to the
quarks and antiquarks in all different ways in the projectile and two reggeons in each target.
Apart from the transverse momenta the projectile impact factor depends on $q_{1-}$ and $q_{2-}$ with $q_{3-}=-q_{1-}-q_{2-}$.
The target impact factors depend on "+" components of $r$.
The central blob is the vertex $\Gamma_{3R\to 4R}$ for the transition of the three incoming reggeons into the 4 outgoing ones with one of the reggeons
not participating in the interaction. Apart from the
transverse momenta it depends on $\qa$,$\qb$, $\ra$ and $\rd$. As before one has $r_1+r_2=-r_3-r_4=e$ with $e_+=0$.

As before integrations over $q_{1-}$, $q_{2-}$ $r_{1+}$ and $r_{4+}$ are factored out. Only the impact factors depend on them. So one can immediately
do these integrations by using energetic variables $(k-q_1)^2$, $(k-q_1-q_2)^2$, $(l+r_1)^2$ and $(l+r_4)^2$ and rotating the Feynman integration contour
around the cut of the impact factors. In this way one obtains the standard impact functions $D_3^{(0)}$ for the projectile (see e.g.~\cite{barew})
and $D^{(0)}$ for the targets
depending only on the corresponding transverse momenta of the attached reggeons,   multiplied by the overall factor   $2s^2/k_+$ entering
(\ref{defh}).
Also since the singularities of the impact factors are supposed to lie at limited values the momenta,
$q_-$, $r_{1+}$ and $r_{4+}$ become  zero at large energies.
As a result the incoming reggeons acquire $q_-=0$ and the outgoing ones $r_+=0$ in accordance with the effective action.
Note that for the reggeon which does not interact one has $q_+=r_-=0$.
In this way one obtains the diffractive amplitude $H_3$ coming from the contributions with three reggeons attached to the projectile
as a sum of  integrals
\beq
H_3^{A,B,C.D}(e)=\int d\tau_\perp D_3^{(0)}(q_3,q_2,q_1)P^{(0)}(r_1,r_2)P^{(0)}(r_4,r_3)\int\frac{dq_+ dr_-}{(2\pi)^2}\Gamma^{A,B,C,D}_{2R\to 4R}(q_+,r_-),
\label{d3}
\eeq
where the integration variables $q_+$ and $r_-$ refer to different momenta in the diagrams A,...D corresponding to  two remaining integrations over
longitudinal momenta.

The integration over $\qc$ in Figs. \ref{fig6},A and C is done due to the cut which provides $-2\pi\delta\Big((q_3+e)^2\Big)$. One gets factor $-1/2|\e|$
and puts $\qc=-p_\perp^2/2\e <0$. The integration over $\qa$ in Figs. \ref{fig6}, B and D is done using $-2\pi\delta\Big((q_a-e)^2\Big)$.
One gets the same factor $-1/2|\e|$ and puts $\qa=p_\perp^2/2\e>0$.
Integrations over $r_4$ in diagrams of Fig. \ref{fig5},A and C give $-1/4|\qc|$  and that over $\ra$ in diagrams B and D give factor
$-1/|\qa|$.

The rest part of $\Gamma_{3R\to 4R}(\qa,\ra,\rb)$ in  diagrams in Fig. \ref{fig5} contains products of vertices $\Gamma_{R\to G+R}$
and integrated $\Gamma_{R\to G+2R}$ namely
\[
H_3^A=\frac{1}{2|\e||\qc|}g^4C_A\int d\tau_\perp D_3^{(0)}(q_3,q_2,r_1)\]\[\times P^{(0)}(r_1,r_2)P^{(0)}(r_4,r_3)\qc B(-p,r_3,r_4)L(p,r_2),\]
\[
H_3^B=-\frac{1}{2|\e||\qa|}g^4C_B\int d\tau_\perp D_3^{(0)}(r_4,q_2,q_1)\]\[\times P^{(0)}(r_1,r_2)P^{(0)}(r_4,r_3)\qa B(p,r_2,r_1)L(-p,r_3),\]
\[
H_3^C=\frac{1}{2|\e||\qc|}g^4C_C\int d\tau_\perp D_3^{(0)}(q_3,r_2,q_1)\]\[\times P^{(0)}(r_1,r_2)P^{(0)}(r_4,r_3)\qc B(-p,r_3,r_4)L(p,r_1),\]
\beq
H_3^D=-\frac{1}{2|\e||\qc|}g^4C_D\int d\tau_\perp D_3^{(0)}(q_3,r_2,q_1)\]\[\times P^{(0)}(r_1,r_2)P^{(0)}(r_4,r_3)\qa B(p,r_2,r_1)L(-p,r_4).
\label{d31}
\eeq
Here $C$'s are the colour factors..
Note that conjugation of $q_3B(-p,3_3,r_4)$ does not change it but conjugation of $L(-p,r_3)$ or $L(-p,r_4)$ changes their sign. Hence an additional
minus in $H_{3}^B$ and $H_3^D$. Now we take into account that $\qa/|\qa|=-\qc/|\qc|=1$, so that the final numerical coefficients are all equal to $-1$
in all $H_3$'s. We use (\ref{derivk}) to express the product $B\otimes L$ summed over polarizations via the Bartels kernel $K_{2\to 3}$. We get
considering all  vectors Euclidean 2-dimensional
\[ B(-p,r_3,r_4)L(p,r_2)=\Big(\frac{-p+r_3}{(-p+r_3)^2}-\frac{q_3}{q_3^2}\Big)\Big(\frac{p}{p^2}-\frac{p+r_2}{(p+r_2)^2}\Big)=
\frac{1}{2}K_{2\to 3}(r_4,r_3,r_2|q_3,q_2),\]
\[ B(p,r_2,r_1)L(-p,r_3)=\Big(\frac{p+r_2}{(p+r_2)^2}-\frac{q_1}{q_1^2}\Big)\Big(-\frac{p}{p^2}+\frac{p-r_3}{(p-r_2)^2}\Big)=
\frac{1}{2}K_{2\to 3}(r_3,r_2,r_1|q_2,q_1),\]
\[ B(-p,r_3,r_4)L(p,r_1)=\Big(\frac{-p+r_3}{(-p+r_3)^2}-\frac{q_3}{q_3^2}\Big)\Big(\frac{p}{p^2}-\frac{p+r_1}{(p+r_1)^2}\Big)=
\frac{1}{2}K_{2\to 3}(r_4,r_3,r_1|q_3,q_1),\]
\[ B(p,r_2,r_1)L(-p,r_4)=\Big(\frac{p+r_2}{(p+r_2)^2}-\frac{q_1}{q_1^2}\Big)\Big(-\frac{p}{p^2}+\frac{p-r_4}{(p-r_4)^2}\Big)=
\frac{1}{2}K_{2\to 3}(r_4,r_2,r_1|q_3,q_1).\]

Now the color factors. The color factor coming from the projectile impact factor is $(-1/2)f^{a_3a_2a_1}$ ~\cite{barew}.
So we get
\[-2C_A=f^{a_3db_3}f^{dcb_3}f^{a_2a_1c}f^{a_3a_2a_1}=-N^4,\]
\[-2C_C=f^{a_3db_3}f^{dcb_3}f^{a_1a_2c}f^{a_3a_2a_1}=N^4,\]
\[-2C_B=f^{a_1b_1d}f^{db_1c}f^{a_2ca_3}f^{a_3a_2a_1}=-N^4,\]
\[-2C_D=f^{a_1b_1d}f^{db_1c}f^{a_3ca_2}f^{a_3a_2a_1}=+N^4.\]

Collecting all factors we finally find
\beq
H_3=H_3^A+H_3^B+H_3^C+H_3^D
\label{d32}
\eeq
where
\[
H_3^A=\frac{1}{8|\e|}g^4N^4\int d\tau_\perp D_3^{(0)}(q_3,q_2,r_1)P^{(0)}(r_1,r_2)P^{(0)}(r_4,r_3)K_{2\to 3}(r_4,r_3,r_2|q_3,q_2),\]
\[
H_3^B=\frac{1}{8|\e|}g^4N^4\int d\tau_\perp D_3^{(0)}(r_4,q_2,q_1)P^{(0)}(r_1,r_2)P^{(0)}(r_4,r_3)K_{2\to 3}(r_3,r_2,r_1|q_2,q_1),\]
\[
H_3^C=-\frac{1}{8|\e|}g^4N^4\int d\tau_\perp D_3^{(0)}(q_3,r_2,q_1)P^{(0)}(r_1,r_2)P^{(0)}(r_4,r_3)K_{2\to 3}(r_4,r_3,r_1|q_3,q_1),\]
\beq
H_3^D=-\frac{1}{8|\e|}g^4C_D\int d\tau_\perp D_3^{(0)}(q_3,r_2,q_1)P^{(0)}(r_1,r_2)P^{(0)}(r_4,r_3)K_{2\to 3}(r_4,r_2,r_1|q_3,q_1).
\label{d33}
\eeq
The found $H_3$ again differs from the one which was obtained in the multiple cut  approach only by factor $1/|\e|$, which
carries the desired $e$-dependence.

\section{Four reggeons attached to the projectile}
With four reggeons attached to the projectile its impact factor depends on three minus components of the longitudinal momenta
$q_{1-}$, $q_{2-}$ and $q_{3-}$ with $q_{4-}=-q_{1-}-q_{2-}-q_{3-}$. As before we use the energetic variables
$(k-q_1)^2$, $(k-q_1-q_2)^2$ and $k-q_1-q_2-q_3)^2$ and rotate the integration contour to enclose the right-hand singularities.
As a result we get the standard impact factor $D_{4}^{(0)}$ found in the multicut approach, which depends only on the transverse momenta
~\cite{barew}.
Also  all minus component of momenta $q_i$, $=1,...4$ are put to zero. The target impact factors are considered as before
all plus components of the outgoing reggeon momenta become equal to zero.

\begin{figure}
\begin{center}
\epsfig{file=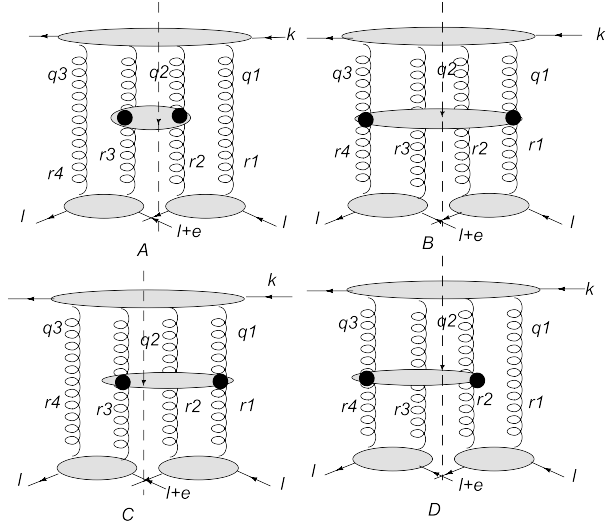, width=8 cm}
\caption{The diffractive amplitudes with four reggeons attached to the projectile}
\label{fig7}
\end{center}
\end{figure}

In the lowest order we have two non-interacting reggeons with all their longitudinal momenta zero.
So they  cannot be coupled to the  same target, unless $\e$ is  zero and contribute only to the low-mass diffraction.
 Thus with $\e\neq 0$ we find
four diagrams for the amplitude shown in Fig. \ref{fig7}

The integration over $\qb$ in Figs.\ref{fig7},A and D is done due to the cut which provides $-2\pi\delta\Big((q_2-e)^2\Big)$. One gets factor $-1/2|\e|$
and puts $\qb=-p_\perp^2/2\e <0$.The integration over $\qa$ in Figs. \ref{fig7}, B and C is done using $-2\pi\delta\Big((q_1-e)^2\Big)$.
One gets the same factor $-1/2|\e|$ and puts $\qa=p_\perp^2/2\e>0$.
There are no additional longitudinal integrations.

At this point it is convenient to sum over colors.
The projectile impact factor contains two pieces with different color factors ~\cite{barew}
\[D_5=-g^2d^{a_4a_3a_2a_1}F_1^{(0)}--g^2d^{a_3a_4a_2a_1}F_2^{(0)},\]
where
\[
F_1^{(0)}=D^{(0)}(1,234)+D^{0}(4,123)-D^{(0)}(14,23),\]
\beq
F_2^{(0)}=D^{(0)}(124,3)+D^{(0)}(134,2)-D^{(0)}(12,34)-D^{(0)}(13,24).
\label{ff}
\eeq
Here
\[d^{abcd}={\rm Tr}(t^at^bt^ct^d+t^dt^ct^bt^a)\]
and we denote $q_1=1$, $q_1+q_2=12$ etc. Summation over colors gives
\[ \frac{1}{4}g^2N^4(F^{(0)}_1-F^{(0)}_2)\ \ {\rm for}\ \ A,B,\ \
 -\frac{1}{4}g^2N^4(F^{(0)}_1-F^{(0)}_2)\ \ {\rm for}\ \ C,D.\]

So we find the amplitude as the transversal integral over the product of two Lipatov vertices together with the
impact factors. Namely
\[H_4^A=-\frac{1}{8|\e|}g^4N^4\int d\tau_\perp F^{(0)}(r_4,q_3,q_2,r_1)L(-p,r_3)L(p,r_2)P^{(0)}(r_4,r_3)P^{(0)}(r_2,r_1),\]
\[H_4^B=-\frac{1}{8|\e|}g^4N^4\int d\tau_\perp F^{(0)}(q_4,r_3,r_2,q_1)L(-p,r_4)L(p,r_1)P^{(0)}(r_4,r_3)P^{(0)}(r_2,r_1),\]
\[H_4^C=\frac{1}{8|\e|}g^4N^4\int d\tau_\perp F^{(0)}(r_4,q_3,r_2,q_1)L(-p,r_3)L(p,r_1)P^{(0)}(r_4,r_3)P^{(0)}(r_2,r_1),\]
\[H_4^D=\frac{1}{8|\e|}g^4N^4\int d\tau_\perp F^{(0)}(q_4,r_3,q_2,r_1)L(-p,r_3)L(p,r_2)P^{(0)}(r_4,r_3)P^{(0)}(r_2,r_1).\]
Here \beq
F^{(0)}=F_1^{(0)}-F_2^{(0)}
\label{deff}
\eeq.
The sign takes into account that conjugation changes the sign of the Lipatov vertex.

Summation over polarization leads to the BFKL kernel $K_{2\to 2}$.
We find (with Euclidean vectors) for diagram in Fig. \ref{fig6}, A.
\[L(-p,r_3)L(p,r_2)=\Big(\frac{-p}{p^2}-\frac{-p+r_3}{(-p+r_3)^2}\Big)\Big(\frac{p}{p^2}-\frac{p+r_2}{p+r_2)^2}\Big)\]\[
=-\frac{1}{p^2}-\frac{(p,-p+r_3)}{p^2(-p+r_3)^2}+\frac{(p,p+r_2)}{p^2(p+r_3=2)^2}+\frac{(-p+r_3,p+r2)}{(-p+r_3)^2(p+r_2)^2}\]\[=
-\frac{1}{p^2}-\frac{1}{2p^2(-p+r_3)^2}\Big(r_3^2-p^2-(-p+r_3)^2\Big)\]\[
-\frac{1}{2p^2(p+r_2)^2}\Big(r_2^2-p^2-(p+r_2)^2\Big)+
\frac{1}{2(-p+r_3)^2(p+r_2)^2}\Big((r_3+r_2)^2-(-p+r_3)^2-(p+r_2)^2\Big)\]\[=
\frac{(r_2+r_3)^2}{2(-p+r_3)^2)(p+r_2)^2}-\frac{r_3^2}{2p^2(-p+r_3)^2}-\frac{r_2^2}{p^2(p+r_2)^2}\]\beq
=\frac{1}{2}\Big(\frac{(r_2+r_3)^2}{q_2^2q_3^2}-\frac{r_3^2}{q_3^2p^2}-\frac{r_2^2}{q_2^2p^2}\Big)=
\frac{1}{2}K_{2\to 2}(r_3,r_2|q_3,q_2).
\eeq
Using the same results for the rest of diagrams we finally find
\[H_4^A=-\frac{1}{16|\e|}g^4N^4\int d\tau_\perp F^{(0)}(r_4,q_3,q_2,r_1)K_{2\to 2}(r_3,r_2|q_3,q_2)P^{(0)}(r_4,r_3)P^{(0)}(r_2,r_1),\]
\[H_4^B=-\frac{1}{16|\e|}g^4N^4\int d\tau_\perp F^{(0)}(q_4,r_3,r_2,q_1)K_{2\to 2}(r_4,r_1|q_4,q_1)P^{(0)}(r_4,r_3)P^{(0)}(r_2,r_1),\]
\[H_4^C=\frac{1}{16|\e|}g^4N^4\int d\tau_\perp F(^{(0)}r_4,q_3,r_2,q_1)K_{2\to 2}(r_3,r_1|q_3,q_1)P^{(0)}(r_4,r_3)P^{(0)}(r_2,r_1),\]
\[H_4^D=\frac{1}{16|\e|}g^4N^4\int d\tau_\perp F^{(0)}(q_4,r_3,q_2,r_1)K_{2\to 2}(r_4,r_2|q_4,q_2)P^{(0)}(r_4,r_3)P^{(0)}(r_2,r_1).\]

Again we observe that the result obtained in the effective action technique differs from the one in the multicut approach only by the factor
$1/|\e|$ which exhibits the dependence on the longitudinal momentum transfer and is missing in the multicut technique.

\section{Evolution, triple pomeron vertices and cross-sections}
\subsection{Evolution. Diffractive vertex}
As we obtained in the effective action approach one obtains the same triple pomeron amplitude as derived in the multicut technique
 but with an extra factor $1/|\e|$, which carries the desired dependence on the longitudinal momentum transfer. The transverse integral is the same
as obtained long ago ~\cite{bartels, barwue}. As a result to study the low-$x$ evolution and express the amplitude
via the standard triple pomeron vertex we can use these old  papers for manipulations in the transverse space (see also later papers  ~\cite{bra,bravac}
where these manipulations are closer to the present ones).

 First of all one notes that in our formulas all initial impact factors in the end are either $D^{(0)}$ as in $H_2$ or its combinations
 depending on different momenta $q_{+}$ or $r_-$ of the  reggeons. This allows to rewrite our results in the form in which the integral starts with $D^{(0)}$
 integrated over its arguments. We get for the whole amplitude (in Euclidean 2 dimensional momenta)
 \beq
 H^{(0)}=\frac{1}{|\e|}\int d\tau D^{(0)}(q,-q)P^{(0)}(r_1,e-r_1)P^{(0)}(r_4,-e-r_4)Z(r_4,r_1|q),
 \label{dd}
 \eeq
 where $d\tau =d^2qd^2r_1d^2r_4(2\pi)^{-9}$ and the so called diffractive triple pomeron vertex $Z(r_4,r_1|q)$ ~\cite{bra,bbv}
 corresponds to  the sum of all transitions from 2,3 and 4 initial reggeons as obtained after passing to integration
 to the "+" momentum in all $D^{(0)}$ in the projectile impact factors.

This expression corresponds to the lowest order in the coupling constant. In the leading log approximation higher orders correspond to introducing
either BFKL interactions between the reggeons coupled to the same $D^{(0)}$ or Regge trajectories into the reggeon propagators.
They describe low $x$ evolution. At this point we can use our old result that this evolution leads to the change of all three $D^{(0)}$ in (\ref{dd})
into the fully evolved $D_y$ which are obtained after evolution in rapidity up to $y$ according to  BFKL  equation. The standard pomeron at rapidity $y$ is just
$D_y(k_1,k_2)/k_1^2k_2^2$. The rapidity $y$ is measured by its value for the real gluon in the cut
\beq
y=\frac{1}{2}\ln \frac{p_+}{p_-}=\frac{1}{2}\ln\frac{p^2}{2|\e|^2}=\ln\frac{p}{|\e|\sqrt{2}}.
\eeq
Momentum $p$  is  in principle determined by the integration variable in (\ref{dd}). In the contribution $H_2$ with only two reggeons
coupled to the projectile we find $p^2=(q-r_1)^2$. However with more reggeons coupled to the projectile this relation is more complicated due to the
passage to the arguments of $D^{(0)}$ in $D_3^{(0)}$ or $D_4^{(0)}$. So rigorously speaking rapidity $y$ is different for different contributions to our amplitudes.
However our derivation is valid for large values of $y$ and the average $p$ is determined by the transverse dimension $R$ of the participant particles
which is finite. So in fact with a logarithmic precision $y=-\ln|\e|$ and can be considered the same for all contributions.

So after evolution at large  rapidities neglecting their variation with $p^2$  at fixed rapidity $y$
we get the amplitude $H^{(HM)}$ corresponding to high-mass diffraction
\beq
H^{(HM)}(\e)=\frac{1}{|\e|}\int d\tau D_{Y-y}(q,-q)P_y(r_1,e-r_1)P_y(r_4,-e-r_4)Z(r_4,r_1|q),
\label{dd1}
\eeq
where $y=-\ln |\e|$ and $Y$ is the overall rapidity of the collision.
Remarkably evolution only changes the incoming and out going pomerons, the diffractive vertex $Z$, which describes their interaction,
remains intact.
Inserting this expression into (\ref{eqa2}) we find the cross-section for high-mass diffractive cross-section
\beq
d\sigma=\frac{dM^2d^2l'_\perp}{8\pi^3M^2}
\int d\tau D_{Y-y}(q,-q)P_y(r_1,e-r_1)P_y(r_4,-e-r_4)Z(r_4,r_1|q).
\label{hmd}
\eeq
This is the same cross-section that was obtained in ~\cite{bartels} long ago. So the only achievement by  using the effective action is a somewhat more rigorous
separation of the amplitude into the interacting pomerons and the vertex describing their interaction, which was in fact
assumed to be valid  in ~\cite{bartels}.

\subsection{Screening correction to the scattering on the deuteron}

\begin{figure}
\begin{center}
\epsfig{file=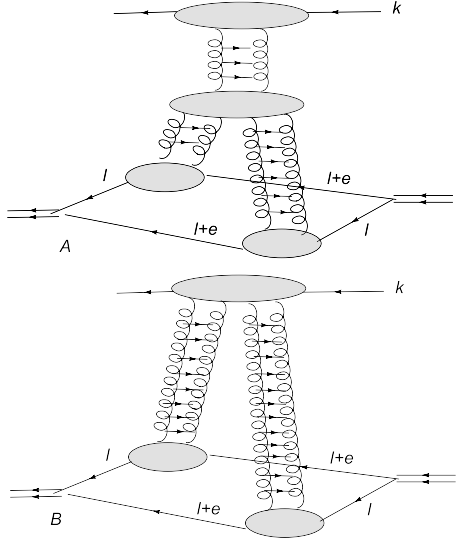, width=6 cm}
\caption{High-mass diffraction (A) and double pomeron exchange (B)}
\label{fig8}
\end{center}
\end{figure}

Passing to the double cross-section on the composite target we first note that after evolution not only diagrams similar  to the high-mass
diffraction , Fig. \ref{fig8},A,  contribute to the cross-section but also ones with actually zero-mass diffraction described by the double
pomeron exchange Fig.\ref{fig8},B and  omitted up to now. So we have to add the double pomeron exchange (DP) contribution.
Unlike our previous calculations for this contribution energy $(k-q_1-q_2)^2=(k-e)^2$ (in Lorenz vectors) is fixed.
To get the standard impact factor one should integrate over $\e$
Actually according to our expressions for the double scattering
(\ref{crseca1}) or (\ref{crsecd1}) we need precisely the integral of the $H^{(DP)}$ over $\e$.  Taking as $(k-e)^2$ in 4-dimensional Lorenz momenta as variable
and closing the contour on the discontinuity of the projectile blob we obtain the impact parameter $D_5^{(0)}$ coupled to two outgoing pomerons.
So we find the DP contribution to the cross-section as
\beq
\int d\e H^{(DP)}(\e)=\int d\tau F^{(0)}(r_4,-r_4,-r_1,r_1)P_Y(r_4,-r_4)P_Y(-r_1,r_1).
\label{dp}
\eeq
To find the total contribution we take into account that $d|\e|/|\e|=dy$ so that the contribution from the high-mass diffraction is
\beq
\int d\e H^{(HM)}(\e)= \int_0^Y dy\int d\tau D_{Y-y}(q,-q)P_y(r_1,-r_1)P_y(r_4,-r_4)Z(r_4,r_1|q).
\label{zp}
\eeq

\begin{figure}
\begin{center}
\epsfig{file=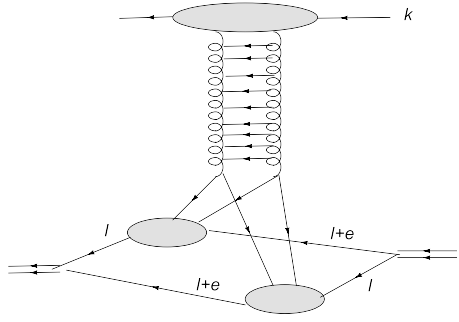, width=6 cm}
\caption{The amplitude with the reggeized piece $D^R$}
\label{fig9}
\end{center}
\end{figure}

As was demonstrated long ago  ~\cite{bartels, bravac, bbv} the sum of DP contribution and that from the diffractive vertex (\ref{zp}
can be rewritten such a way that the double pomeron exchange is eliminated and instead a single pomeron exchange appears together with the standard
triple pomeron vertex $V$:
\beq
\int d\e(H^{(DP)}+H^{(HM)})=\int d\e(H^{(R)}+H^{(3P)}).
\label{split}
\eeq
The two parts $H^{(R)}$ and $H^{(3P)}$ behave differently at high energies:
The so-called reggeized piece $H^{(R)}$  is shown in Fig. \ref{fig9}.  Integrated over $\e$ it is just the impact factor $F^{(0)}$  for the four reggeons attached to the projectile in
which all functions $D^{(0)}$ are substituted by their evolved expressions $D_Y$ with all their arguments retained. The second part $D^{(3P)}$ is given by
(\ref{zp}) in which the diffractive vertex $Z$ is substituted by the standard triple pomeron vertex $V$.
The latter can be conveniently written in the coordinate space ~\cite{bravac}
\beq
\Gamma^{3P}(x_1,x_2,x_3)=-\frac{g^4N}{4\pi^3}\frac{(x_1-x_3)^2\nabla_1^2\nabla_3^2}
{(x_1-x_2)^2(x_2-x_3)^2},
\label{vdef}
\eeq
where each pomeron is assumed to contain factor $N$.

Using (\ref{split}) we find the cross-section for the double scattering on the deuteron as a sum of two terms
\beq
\sigma_d^{double}=\sigma_d^{(R)}+\sigma_d^{(3P)},
\label{crsecd2}
\eeq
where
\beq
\sigma_d^{(R)}=-\frac{1}{2\pi^3}\Big<\frac{1}{r^2}\Big>_d\int\frac{d^2r_1d^2r_4}{(8\pi^3)^2}  F_Y(r_4,-r_4,-r_1,r_1)P^{(0)}(r_1,-r_1)P^{(0)}(-r_4,r_4)
\label{crsecdr}
\eeq
and
\beq
\sigma_d^{(3P)}=-\frac{1}{\pi^2}\Big<\frac{1}{r^2}\Big>_d
\int_0^Ydy \int \frac{d^2qd^2r_1d^2r_4}{(8\pi^3)^3} D_{Y-y}(q,-q)P_y(r_1,-r_1)P_y(-r_4,r_4)V(r_4,r_1|q).
\label{crsecd3p}
\eeq
Here $F_Y=F^{(0)}(D^{0}\to D_Y)$.
Both $\sigma_d^{(R)}$ and $\sigma_d^{(3P)}$ turn out to be negative.
Their sum gives the so-called screening correction to the main part of the
cross-section given by the sum of the cross-sections on the proton and on the neutron.
In the BFKL approach the latter is just the sum of single pomeron exchanges for the proton and for the neutron.
The advantage of splitting the cross-section into parts $\sigma^{(R)}$ and $\sigma^{(3P)}$ lies in their different behavior at large energies:
part $\sigma^{(R)}$ grows as a single pomeron and part $\sigma^{(3P)}$ grows twice rapidly, as two pomerons.

The double cross-section on the nucleus is obtained from (\ref{crseca1}) in a similar manner.

The reggeized term (\ref{crsecdr}) can be simplified when we use the explicit form of $F_y$. From (\ref{deff}) we see that
 $F_y(r_4,-r_4,-r_1,r_1)$ contains  terms which depend only on $r_1$, only on $r_4$, terms which  depend neither on $r_1$ nor on $r_2$  and
 finally terms which depend on  $r_1\pm r_4$. The first three groups do not give contribution, since according to color transparency
$P_Y(r,-r)$ integrated over its momentum gives zero. In the coordinate space it describes the dipole of zero dimension. Note that this property
is valid at all values of $Y$ and at $Y=0$ in particular. We assume that $P^{(0)}(r,-r)$ can be taken as the limit of $P_Y(r,-r)$ at $Y\to 0$.
This follows from its expression via the Green function in which the coupling constant enters only through the combination $YE$ where $E$ are the BFKL levels, which vanishes when
$Y\to 0$ or $g\to 0$. Obviously this  is a regularization of the expression for $P^{(0)}(r,-r)=D^{(0)}(r,-r)/r^4$.
Thus the only terms in $F_y$  (which give the same contribution)
are $-D_Y(r_1+r_4,-r_1-r_4)$ and $-D_Y(r_1-r_4,-r_1+r_4)$. So writing $D_Y^(q,-q)$ simply as $D_Y(q)$ we get the reggeized part as
\beq
\sigma_d^{(R)}=\frac{1}{2\pi^3}\Big<\frac{1}{r^2}\Big>_d\int\frac{d^2r_1d^2r_4}{(8\pi^3)^2} D^{(0)}(r_1)D^{(0)}(r_4)\frac{D_Y(r_1+r_4)}{r_1^4r_4^4}.
\label{crsecdr1}
\eeq
In this form convergence at small $r_1$ and $r_4$ is not obvious. $D^{(0)}$ vanishes at $r\to 0$ at least as $r^2$ but this may lead to the logarithmic divergence.
However one can subtract from $D_Y(r_1+r_4)$ its values at $r_1=0$ and $r_3=0$ without changing the result
\beq
\sigma_d^{(R)}=\frac{1}{2\pi^3}\Big<\frac{1}{r^2}\Big>_d\int\frac{d^2r_1d_2r_4}{(8\pi^3)^2} D^{(0)}(r_1)D^{(0)}(r_4)\frac{D_Y(r_1+r_4)-D_Y(r_1)-D_Y(r_4)}{r_1^4r_4^4}.
\label{crsecdr2}
\eeq
Now the numerator vanishes when either $r_1$ or $r_4$ are equal to zero, which provides convergence at $k_1=0$ or $k_4=0$.  The form (\ref{crsecdr2}) for the reggeized part
can be used for practical calculations.

The total cross-section on the deuteron is thus
\beq
\sigma^{tot}_d=\sigma_p+\sigma_n-\sigma^{screen}\simeq 2\sigma_p-\sigma^{screen},
\label{sigtot}
\eeq
where $\sigma^{screen}$ is the sum (\ref{crsecd2}) with the opposite sign, which turns out to be positive.

It is instructive to compare the two components of the screening correction for a more or less realistic situation.
Qualitative estimate show that
\[ \sigma_d^{(R)}(Y)\sim \bal^2e^{\Delta Y},\ \ \sigma_d^{(3P)}(Y)\sim \bal^2e^{2\Delta Y},\]
where
$\bal=\alpha_sN/\pi$ and $\Delta=4\bal \ln 2$.
Also both contain the small factor $<1/r^2>_d\simeq 0.48\ fm^{-2}$ (with the Hulthen wave function).
So with a small coupling constant and finite $Y$ both terms in the screening correction are small
compared to the main part $2\sigma_p$.
With the growth of $Y$ the ratio $\sigma_d^{(R)}/2\sigma_p$ remains intact whereas
the ratio $\sigma_d^{(3P)}/2\sigma_p$ grows as $\exp(\Delta Y)$. So at sufficiently high $Y$
this ratio becomes greater than unity and the total cross-section becomes negative. Obviously such values of $Y$
lie outside the applicability of the BFKL approach in the leading log approximation.

To make more quantitative estimations we choose the coupling constant to have the intercept $\Delta$ more or less in accordance with the
observed growth  of $\sigma_p(Y)$ at large $Y$. We choose $\Delta=0.12$ which fixes the coupling constant to be quite small
$\bal=0.0432$. In calculating  $\sigma_d^{(3P)}$ we encounter a problem with the infrared behavior. Actually we do not find any
infrared divergence. However the important values of momenta at large rapidities shift deep into the infrared region making the contribution abnormally large
and providing an additional growth with $Y$. So one has to impose the confinement and restrict small values of momenta to lie
in the physically reasonable interval $q>q_{min}\sim \Lambda_{QCD}$. Taking the projectile to be the proton and perform calculations one has to choose
the form of the proton color density. We take it to be
\beq
\Phi(q)=q^2\rho(q),\ \ \rho(q)=\gamma e^{-\beta q^2},
\label{rho}
\eeq
where $\beta$ is dictated by the proton radius $0.08\ fm$ and $\gamma$ is determined from the proton cross-section at small $Y$ which we take $4\ fm^2$.

\begin{figure}
\begin{center}
\epsfig{file=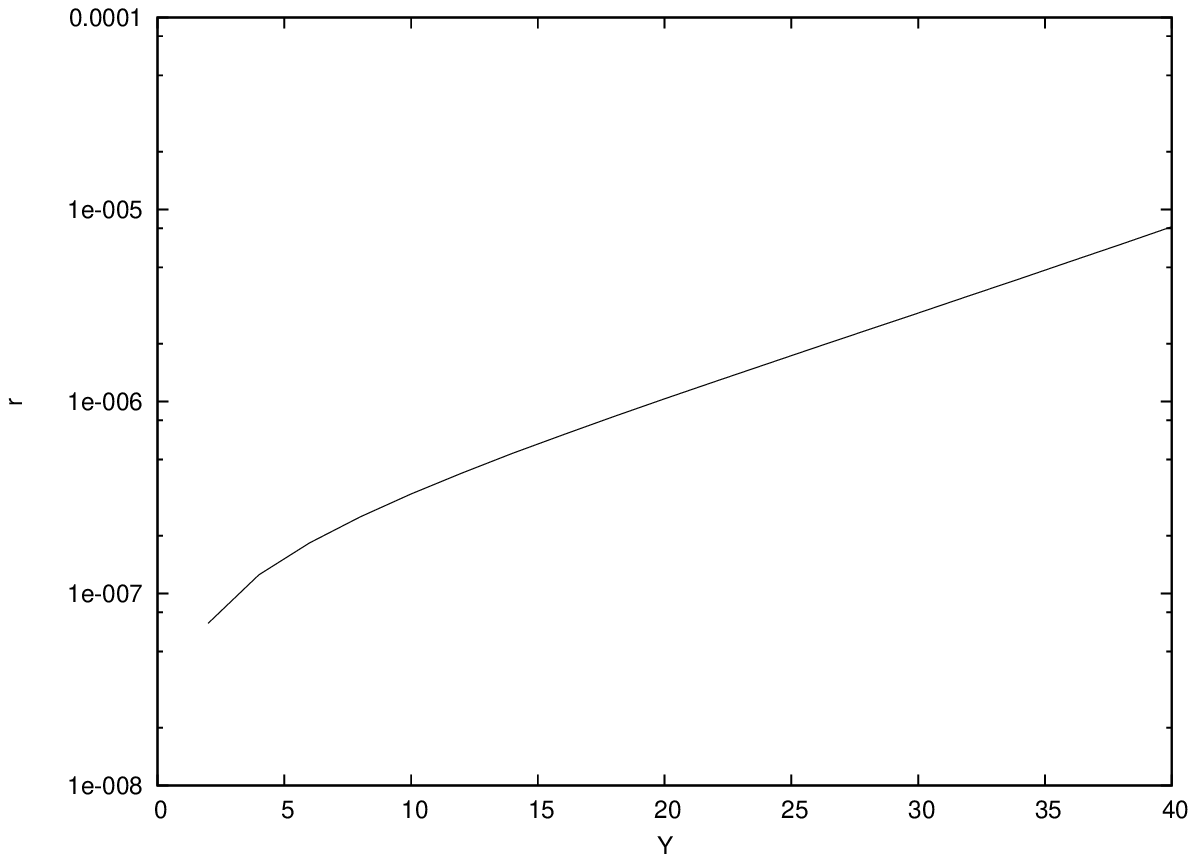, width=6 cm}
\epsfig{file=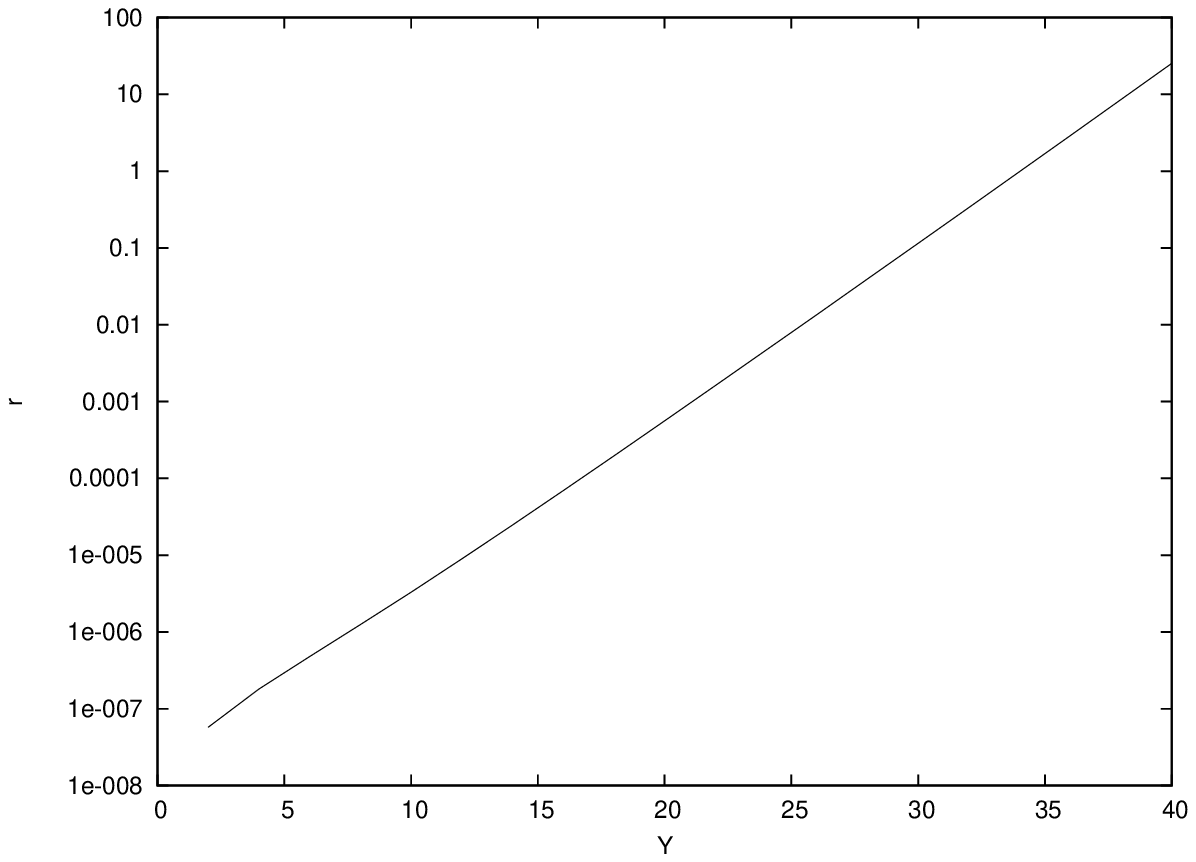, width=6 cm}
\caption{The ratio of the triple pomeron part to the main one with the opposite sign
with $\bal=0.0432$ (the left panel) and with $\bal=0.2$ (the right panel)}
\label{fig10}
\end{center}
\end{figure}

Relegating  some details of the calculation to Appendix we only present here our results. We assume $Y$ to be large so that one can use the asymptotic
expressions for the relevant pomeron functions.
As expected the ratio of the reggeized part to the main one does not depend on $Y$ and
\beq
-\frac{\sigma^{(R)}(Y)}{2\sigma_p(Y)}=0.00566.
\eeq
The ratio of the triple pomeron part to the main one turns out to be still much smaller. But it steadily grows with $Y$. We
present $r(Y)=-\sigma^{(3P)}(Y)/2\sigma_p(Y)$ in Fig. \ref{fig10} in the left panel.

Just for illustration we also consider the case of a much larger coupling constant  $\bal=0.2$ frequently
used in many older calculations. With this choice
\beq
-\frac{\sigma^{(R)}(Y)}{2\sigma_p(Y)}=0.121
\eeq
and  ratio $r(Y)$ is shown in Fig. \ref{fig10} in the right panel

\section{Conclusions}
We have studied the triple-pomeron amplitude in the effective action formalism with the aim of deriving
its dependence on the transferred longitudinal momentum, necessary for the calculation of cross-sections.
We limited ourselves to the imaginary part of the amplitude, which substantially simplified our task.
Our results turned out quite simple: the dependence on the longitudinal momentum transfer is separated into a simple extra factor.
Using old studies performed in the multiple cut approach we transformed our amplitudes and the resulting cross-sections into
the more or less standard forms where either the single or double  pomeron exchange appear accompanied by normal or diffractive
triple pomeron vertices respectively. The found high-mass diffractive cross-section off a hadron coincides with the known one,
On the other hand our results allow to obtain a rigorous expression for the double scattering cross-sections on a composite target.
Estimation indicate that at present energies the screening corrections corresponding to the double cross-section are dominated by
 the so-called reggeized contribution, which however is still much smaller
tan the bulk given by the sum of the cross-sections on the constituents

In principle the effective action approach allows to calculate also the real part of the amplitudes. This is necessary for instance to find the elastic
cross-sections off the composite target. Unfortunately such calculations turn out to be much more complicated due to the necessity to use a general gauge
and struggle against appearing divergencies. We retain our hope to advance in this direction.

\section{Appendix. Some details on the calculation of $\sigma_d^{(3P)}$ and $\sigma_d^{(R)}$}
\subsection{BFKL details}
As a basis we use the leading  semi-amputated (SA) eigenfunctions of the BFKL equation
in the forward direction
\beq
\phi_{n=0\nu}(k,\chi)=\sqrt{2}k^{-1+2i\nu}, \ \ -\infty <\nu<+\infty.
\label{phi}
\eeq
The full pomeron eigenfunction is $\Phi_{0\nu}(k)=\phi_{0\nu}(k)/k^2$.
The corresponding SA Green function is
\beq
g_y(k,k')=\int d\nu e^{y\omega_{0\nu}}\phi_{0\nu}(k)\phi_{0\nu}^*(k'),
\label{sag}
\eeq
where the gluon trajectories $\omega_{0\nu}$ are the known eigenvalues of the BFKL equation.
At $y=0$ we have
\beq
g_0(k,k')=\frac{2\pi}{k}\delta(k-k')
\eeq
and at large $y$
\beq
g^{as}_y(k,k')=\frac{2}{kk'}e^{y\Delta}\sqrt{\frac{\pi}{ay}}\exp\Big(-\frac{\ln^2(k^2/{k'}^2)}{4ay}\Big),
\label{gas}
\eeq
where $a=14\bal\zeta(3)$.

For the pp scattering with the proton color density (\ref{rho}) we find at $y=0$
\beq
\sigma_p(Y=0)\equiv \sigma_0=\int_0^\infty \frac{dk^2}{2\pi}\rho^2(k)=\frac{\gamma^2}{4\pi\beta}
\eeq
and at large $y$
\beq
\sigma^{as}_p=e^{y\Delta}\sqrt{\frac{\pi}{ay}}\Big(\int_0^\infty \frac{dk}{2\pi}\rho(k)\Big)^2=\sigma_0e^{y\Delta}\sqrt{\frac{\pi}{ay}}.
\label{sigas}
\eeq
The SA pomeron coupled to the color density (\ref{rho}) at high energies is given by
\beq
\phi^{as}(q)=\frac{2}{q}e^{y\Delta}\sqrt{\frac{\pi}{ay}}\int\frac{d^2k}{(2\pi )^2k}\rho(k)\exp\Big(-\frac{\ln^2(q^2/k^2)}{4ay}\Big).
\label{phias}
\eeq

\subsection{The triple pomeron contribution}
The bulk of the contribution $\sigma_d^{(3P)}$ given by Eq. (\ref{crsecd3p}) was studied in our paper ~\cite{bra2017}
devoted to the diffractive cross-section off the deuteron. There the integral over momenta was transformed to the coordinate space:
\beq
\sigma_d^{(3P)}=\frac{1}{32\pi^8}g^4N^4\Big<\frac{1}{r^2}\Big>_d
\int_0^Ydy
 \int \frac{d^2x_{12}d^2x_{23}}{x_{12}^2x_{23}^2x_{31}^2} D_{Y-y}(x_{31})P_y(x_{12})P_y(x_{23})
\label{crsecd3p1}
\eeq
with $x_{12}+x_{23}+x_{31}=0$.
The sign takes into account that the triple pomeron vertex $V$ bears the minus sign.
 With the help of $\delta^2(x_{12}+x_{23}+_{31})$
 the integral over coordinates in (\ref{crsecd3p1}) transforms into
\beq
J(y)=\int \frac{d^2q}{(2\pi)^2}\psi_y^2(q)\chi_{Y-y}(q),
\label{defj}
\eeq
where
\beq
\psi_y(q)=\int\frac{d^2x}{x^2}P_y(x)e^{iqx},\ \ \chi_y(q)=\nabla^2q^4\nabla^2\psi_y(q).
\label{defpsi}
\eeq

Using the relation
\beq
q^2\nabla^2\psi_y(q)=-q^2P_y(q)=-\phi_y(q)
\label{relation}
\eeq
one finds that the coefficients in the expansion of $\psi(q)$ in $\phi_{0n}$
are $-(1-2i\nu)^2$ smaller than of $\phi(q)$ and those of $\chi(q)$ are
$-(1+2i\nu)^2$ larger the coefficients of $\phi(q)$.
It follows that asymptotically at high $y$ we have
$\psi(q)^{as}=\chi(q)^{as}=-\phi(q)^{as}$.

Note that in the integral $J(y)$ (\ref{defj}) the product  of asymptotical $\psi^2\chi$ generates a singularity $1/q^3$ at
small $q$. As mentioned, this does not lead to divergence due to the exponential factor. However this factor begins to play its role only
at extremely small $q$ when $\ln(1/q)\sim \sqrt{y}$. Numerical estimates show that this leads to enormous values of the $J$
absolutely beyond any sensible order. So the triple pomeron contribution turns out to be decisively dependent on the infrared
region of momenta or equivalently on large distances. Any reasonable calculation therefore has to limit values of $q$ where
the BFKL approach may be reasonable. We assume that this limitation should restrict $q$ to values above $\Lambda_{QCD}\simeq 0.3$ GeV/c.
To factorize the 8-dimensional integral $J$ and still retain the (part of) behavior of the exponential factor in  (\ref{phias})
we substitute  $\ln(k^2)$ in it by $\ln(1/\beta)$ having in mind good convergence in $k$ with the typical value $k^2\sim 1/\beta$.
Using asymptotical expressions for $\psi$ and $\chi$ we find
\beq
\sigma_d^{(3P)}=-\bal^2N^2C_1\Big<\frac{1}{r^2}\Big>_d
\int_0^Ydye^{\delta(Y+y)}\frac{1}{y\sqrt{Y-y}} I_1(y).
\label{crsecd3p3}
\eeq
where
\[
C_1=\frac{1}{(2\pi)^6}\,\Big(\frac{\gamma^2}{\beta a}\Big)^{3/2}\]
and
\beq
I_1(y)=\int_{q_{min}}^\infty \frac{dq}{q^2}\exp \Big[-\frac{\ln^2(q^2\beta)}{4a}\Big(\frac{2}{y}+\frac{1}{Y-y}\Big)\Big].
\label{defi}
\eeq
The lower limit is
$q_{\min}= 1.5\ fm^{-1}$.
The integral over $y$ is convergent at $y=0$, since $I(y)$ goes to zero at this point. It was calculated numerically.

\subsection {The reggeized contribution}
The reggeized contribution is given by expression
\beq
\sigma_d^{(R)}=\frac{1}{8\pi^5}\Big<\frac{1}{r^2}\Big>_d J_R,
\label{crsecdr3}
\eeq
where
\beq
J_R=\int\frac{d^2r_1d^2r_4}{(2\pi)^4} P^{(0)}(r_1)P^{(0)}(r_4)D_Y(r_1+r_4).
\label{defjr}
\eeq

From (\ref{phias}) we find the asymptotic of $D_Y(q)=q^2\phi(q)$.
This together with the expression for the initial pomeron $P^{(0)}(q)=\rho(q)/q^2$ gives
\beq
J_R=\frac{2}{(2\pi)^3}e^{Y\Delta}\sqrt{\frac{\pi}{aY}}\int \frac{dr_1dr_2d\phi}{r_1r_4}\rho(r_1)\rho(r_4)F\sqrt{r_1^2+r_4^2+2r_1r_4\cos \phi}.
\label{defjr3}
\eeq
To improve convergence we act as in (\ref{crsecdr2}) and subtract from the square root its values at $r_1=0$ and $r_4=0$.
After that we can drop the exponential factor in the integration over $k$, which does not spoil convergence at small $r_1$
and $r_4$ and is certainly possible at very high $Y$.
With $\rho(k)$ given by (\ref{rho}) we then find
\beq
F=\frac{\gamma}{4\pi}\gamma\sqrt{\frac{\pi}{\beta}}.
\eeq
So we have
\beq
J_R=\frac{\gamma^3}{16\pi^3}e^{Y\Delta}\sqrt{\frac{1}{a\beta Y}}I_2,
\label{defr4}
\eeq
where
\beq
I_2=
\int \frac{dr_1dr_4d\phi}{r_1r_4}e^{-\beta(r_1^2+r_4^2)}(\sqrt{r_1^2+r_4^2+2r_1r_4\cos \phi}-r_1-r_4).
\label{defi2}
\eeq
Finally we obtain
\beq
\sigma_d^{(R)}=\bal^2 N^2C_2\Big<\frac{1}{r^2}\Big>_d e^{Y\Delta}\sqrt{\frac{1}{Y}}I_2,
\label{crsecdr4}
\eeq
where
\beq
C_2=\frac{1}{4\pi^3}\gamma^3\frac{1}{\sqrt{\beta a}}.
\eeq
The three-dimensional integral (\ref{defi2}) was calculated numerically.


\begin{thebibliography}{100}
%
\bibitem{bartels} J.Bartels, Z.Phys. {\bf C 60} (1993) 471
%
\bibitem{barwue} J.Bartels and M.Wuesthoff, Z.Phys. {\bf C 66} (1995) 157
%
\bibitem{mueller}A.H.Mueller, Nucl. Phys. {\bf B 415} (1994) 373;
{\bf B 437} (1995) 107; A.H.Mueller and B.Patel, Nucl. Phys. {\bf B 425} (1994) 471

\bibitem{bra5} M.A.Braun, Eur. Phys. J {\bf C 73} (2013) 2418
%
\bibitem{agk} V.A.Abramovsky, V.N.Gribov, O.V.Kancheli, Sov. J. Nucl. Phys. {\bf 18} (1974) 308.
%
\bibitem{barew} J.Bartels, C.Ewerz, JHEP {\bf 9909} (1999) 026
%
\bibitem{hent} M.Hentschinski, {\it dissertation} (2009)  arXiv:0908.2576 [hep-ph]

\bibitem{lipatov} L.N.Lipatov, Nucl. Phys. {\bf B 452} (1995) 369; Phys. Rep., {\bf 286} (1997) 131
%
\bibitem{antonov} E.N.Antonov, I.O.Cherednikov, E.A.Kuraev, L.N.Lipatov, Nucl. Phys. {\bf B 721} (2005) 111.
%
\bibitem{bfkl} L.N.Lipatov,Sov. J.Nucl.Phys. {\bf 23} (1976) 338;
E.A.Kuraev, L.N.Lipatov, V.S.fadin, Sov.Phys.JETP {\bf 45} (1977) 199;
I.I.Balitsky, L.N.Lipatov, Sov. J.Nucl.Phys. {\bf 28} (1978) 822
%
\bibitem{bravyaz} M.A.Braun, M.I.Vyazovsky, Eur. Phys. J. {\bf C 51} (2007) 103
%
\bibitem{bra} M.Braun, Eur. Phys. J., {\bf C 6} (1999) 321.
%
\bibitem{bravac} M.A. Braun and G.P.Vacca, Eur. Phys. J. {C 6} (1997) 147
%
\bibitem{bbv} J.Bartels, M.Braun, G.P.Vacca Eur. Phys. J. {\bf 40} (2005) 419.
%
\bibitem{bra2017} M.A.Braun, Eur. phys. J. {\bf 77} (2017) \#5:279
%
\end{thebibliography}
\end{document}